\PassOptionsToPackage{unicode}{hyperref}
\PassOptionsToPackage{hyphens}{url}
\PassOptionsToPackage{dvipsnames,svgnames,x11names}{xcolor}
\documentclass[
  12pt]{article}

\usepackage{amsmath,amssymb}
\usepackage{iftex}
\ifPDFTeX
  \usepackage[T1]{fontenc}
  \usepackage[utf8]{inputenc}
  \usepackage{textcomp} % provide euro and other symbols
\else % if luatex or xetex
  \usepackage{unicode-math}
  \defaultfontfeatures{Scale=MatchLowercase}
  \defaultfontfeatures[\rmfamily]{Ligatures=TeX,Scale=1}
\fi
\usepackage{lmodern}
\ifPDFTeX\else  
\fi
\IfFileExists{upquote.sty}{\usepackage{upquote}}{}
\IfFileExists{microtype.sty}{% use microtype if available
  \usepackage[]{microtype}
  \UseMicrotypeSet[protrusion]{basicmath} % disable protrusion for tt fonts
}{}
\makeatletter
\@ifundefined{KOMAClassName}{% if non-KOMA class
  \IfFileExists{parskip.sty}{%
    \usepackage{parskip}
  }{% else
    \setlength{\parindent}{0pt}
    \setlength{\parskip}{6pt plus 2pt minus 1pt}}
}{% if KOMA class
  \KOMAoptions{parskip=half}}
\makeatother
\usepackage{xcolor}
\makeatletter
\ifx\paragraph\undefined\else
  \let\oldparagraph\paragraph
  \renewcommand{\paragraph}{
    \@ifstar
      \xxxParagraphStar
      \xxxParagraphNoStar
  }
  \newcommand{\xxxParagraphStar}[1]{\oldparagraph*{#1}\mbox{}}
  \newcommand{\xxxParagraphNoStar}[1]{\oldparagraph{#1}\mbox{}}
\fi
\ifx\subparagraph\undefined\else
  \let\oldsubparagraph\subparagraph
  \renewcommand{\subparagraph}{
    \@ifstar
      \xxxSubParagraphStar
      \xxxSubParagraphNoStar
  }
  \newcommand{\xxxSubParagraphStar}[1]{\oldsubparagraph*{#1}\mbox{}}
  \newcommand{\xxxSubParagraphNoStar}[1]{\oldsubparagraph{#1}\mbox{}}
\fi
\makeatother

\providecommand{\tightlist}{%
  \setlength{\itemsep}{0pt}\setlength{\parskip}{0pt}}\usepackage{longtable,booktabs,array}
\usepackage{calc} % for calculating minipage widths
\usepackage{etoolbox}
\makeatletter
\patchcmd\longtable{\par}{\if@noskipsec\mbox{}\fi\par}{}{}
\makeatother
\IfFileExists{footnotehyper.sty}{\usepackage{footnotehyper}}{\usepackage{footnote}}
\makesavenoteenv{longtable}
\usepackage{graphicx}
\makeatletter
\newsavebox\pandoc@box
\newcommand*\pandocbounded[1]{% scales image to fit in text height/width
  \sbox\pandoc@box{#1}%
  \Gscale@div\@tempa{\textheight}{\dimexpr\ht\pandoc@box+\dp\pandoc@box\relax}%
  \Gscale@div\@tempb{\linewidth}{\wd\pandoc@box}%
  \ifdim\@tempb\p@<\@tempa\p@\let\@tempa\@tempb\fi% select the smaller of both
  \ifdim\@tempa\p@<\p@\scalebox{\@tempa}{\usebox\pandoc@box}%
  \else\usebox{\pandoc@box}%
  \fi%
}
\def\fps@figure{htbp}
\makeatother

\usepackage{booktabs}
\usepackage{longtable}
\usepackage{array}
\usepackage{multirow}
\usepackage{wrapfig}
\usepackage{float}
\usepackage{colortbl}
\usepackage{pdflscape}
\usepackage{tabu}
\usepackage{threeparttable}
\usepackage{threeparttablex}
\usepackage[normalem]{ulem}
\usepackage{makecell}
\usepackage{xcolor}
\makeatletter
\@ifpackageloaded{caption}{}{\usepackage{caption}}
\AtBeginDocument{%
\ifdefined\contentsname
  \renewcommand*\contentsname{Table of contents}
\else
  \newcommand\contentsname{Table of contents}
\fi
\ifdefined\listfigurename
  \renewcommand*\listfigurename{List of Figures}
\else
  \newcommand\listfigurename{List of Figures}
\fi
\ifdefined\listtablename
  \renewcommand*\listtablename{List of Tables}
\else
  \newcommand\listtablename{List of Tables}
\fi
\ifdefined\figurename
  \renewcommand*\figurename{Figure}
\else
  \newcommand\figurename{Figure}
\fi
\ifdefined\tablename
  \renewcommand*\tablename{Table}
\else
  \newcommand\tablename{Table}
\fi
}
\@ifpackageloaded{float}{}{\usepackage{float}}
\floatstyle{ruled}
\@ifundefined{c@chapter}{\newfloat{codelisting}{h}{lop}}{\newfloat{codelisting}{h}{lop}[chapter]}
\floatname{codelisting}{Listing}

\makeatother
\makeatletter
\@ifpackageloaded{caption}{}{\usepackage{caption}}
\@ifpackageloaded{subcaption}{}{\usepackage{subcaption}}
\makeatother

\usepackage[]{natbib}
\usepackage{bookmark}

\IfFileExists{xurl.sty}{\usepackage{xurl}}{} % add URL line breaks if available
\hypersetup{
  pdftitle={A Systematic Literature Review of Undergraduate Data Science Education Research},
  pdfauthor={Mine Dogucu; Sinem Demirci; Harry Bendekgey; Federica Zoe Ricci; Catalina M. Medina},
  pdfkeywords={data science curriculum, data science programs, data
science courses, educational technology, open access},
  colorlinks=true,
  linkcolor={blue},
  filecolor={Maroon},
  citecolor={Blue},
  urlcolor={Blue},
  pdfcreator={LaTeX via pandoc}}

\begin{document}

\def\spacingset#1{\renewcommand{\baselinestretch}%
{#1}\small\normalsize} \spacingset{1}

%%%%%%%%%%%%%%%%%%%%%%%%%%%%%%%%%%%%%%%%%%%%%%%%%%%%%%%%%%%%%%%%%%%%%%%%%%%%%%

\date{January 3, 2025}
\title{\bf A Systematic Literature Review of Undergraduate Data Science
Education Research}
\author{
Mine Dogucu\thanks{Dogucu has been supported by NSF IIS award \#2123366.
Dogucu completed an earlier part of this work in Department of
Statistical Science at University College London.}\\
Department of Statistics, University of California, Irvine\\
and\\Sinem Demirci\thanks{Demirci has been supported by the Scientific
and Technological Research Council of Türkiye. Demirci completed an
earlier part of this work in Department of Statistical Science at
University College London.}\\
Statistics Department, California Polytechnic State University\\
and\\Harry Bendekgey\thanks{Bendekgey has been supported by the HPI
Research Center in Machine Learning and Data Science at UC Irvine.}\\
Department of Computer Science, University of California, Irvine\\
and\\Federica Zoe Ricci\thanks{Ricci has been supported by the HPI
Research Center in Machine Learning and Data Science at UC Irvine.}\\
Department of Statistics, University of California, Irvine\\
and\\Catalina M. Medina\thanks{Medina has been supported by NSF IIS
award \#2123366.}\\
Department of Statistics, University of California, Irvine\\
}
\maketitle

\bigskip
\bigskip
\begin{abstract}
The presence of data science has been profound in the scientific
community in almost every discipline. An important part of the data
science education expansion has been at the undergraduate level. We
conducted a systematic literature review to (1) portray current evidence
and knowledge gaps in self-proclaimed undergraduate data science
education research and (2) inform policymakers and the data science
education community about what educators may encounter when searching
for literature using the general keyword `data science education.' While
open-access publications that target a broader audience of data science
educators and include multiple examples of data science programs and
courses are a strength, significant knowledge gaps remain. The
undergraduate data science literature that we identified often lacks
empirical data, research questions and reproducibility. Certain
disciplines are less visible. We recommend that we should (1) cherish
data science as an interdisciplinary field; (2) adopt a consistent set
of keywords/terminology to ensure data science education literature is
easily identifiable; (3) prioritize investments in empirical studies.
\end{abstract}

\noindent%
\vfill

\newpage
\spacingset{1.9} % DON'T change the spacing!

\newcommand{\github}[1]{https://github.com/mdogucu/comp-data-sci}
\newcommand{\osf}[1]{https://osf.io/b3u7y/}
\newcommand{\ecots}[1]{Electronic Conference on Teaching Statistics 2024}

\section{Introduction}\label{sec-introduction}

The emergence of data science in the last few decades has given
scientists a lot to talk and write about, accumulating a wealth of
scientific literature. The presence of data science has been profound in
the scientific community in almost every discipline. The National
Science Foundation (NSF) called ``Harnessing the Data Revolution'' one
of its 10 big ideas which cuts across all NSF Directorates
\citeyearpar{nsf}.

The demand for data science skills in academia and industry resulted in
many higher education institutions developing courses
\citep[e.g.,][]{baumer} as well as degree programs
\citep[e.g.,][]{glantz, adhikari, stern2021data}. With newer educational
opportunities, newer educational research questions have been formed and
thus data science education emerged as a field. For instance, the
American Statistical Association's (ASA's), \emph{Journal of Statistics
Education} changed its name to \emph{Journal of Statistics and Data
Science Education (JSDSE)} \citep{asajour2020}.

Over the years, a major focus of the data science education community,
unsurprisingly, has been at the undergraduate level. Different
professional organizations and groups have tried to describe the data
science competencies including but not limited to the Park City Math
Institute report \citep{de2017curriculum}, the framework of the National
Academies of Sciences, Engineering, and Medicine
\citeyearpar{nationalacademiesofsciences2018}, the computing
competencies guidelines of the Association for Computing Machinery's
\citep{danyluk2021a}, the accreditation for data science programs of the
Accreditation Board for Engineering and Technology (ABET)
\citeyearpar{abet}, and the EDISON project of the European Union
\citep{wiktorski2017}. It is also worth noting that the majority of data
science programs are at the master's level and there is still a lot of
room for growth at the undergraduate level \citep{li2021}. Given the
global need for data science education, it is important for us to
understand undergraduate data science education as well as the
scientific literature on this topic.

Even though there is not a consensus on what data science is and which
disciplines are part of it, a common view is that data science is
interdisciplinary and statistics, computer science, mathematics and
other domains contribute to it \citep{donoho2017}. The interdisciplinary
structure of data science naturally gets reflected in educational
research as well, making data science education an interdisciplinary
field with contributors from statistics education, computer science
education, and other educational research communities.

While contributions from different educational research communities can
make data science education richer, these contributions may not easily
circulate across disciplines in the absence of a centralized community.
For instance, Hazzan and Mike \citeyearpar{hazzan2021} mention that,
despite its name, JSDSE mainly caters to the statistics community and
state that ``no journal exists today that deals exclusively with data
science education, let alone highlights data science education from an
interdisciplinary perspective''.

The aforementioned examples of courses, programs and curricular
guidelines as well as the existence of conferences and journals with the
title or keywords \emph{data science education} show the growth of this
field. In this manuscript, we detail a study conducted to understand
undergraduate data science education through in-depth readings of the
existing literature. As undergraduate teacher-scholars, our focus was on
data science education at the undergraduate level. In this study, we did
not intend to define data science education research or judge whether a
study meets a specific definition of data science. Instead, we focused
on publications that claimed to be on ``data science education''.

Our goals were to (1) specify current evidence and knowledge gaps in
self-proclaimed undergraduate data science education and (2) inform
policymakers and the data science education community about what
educators may encounter when searching for literature using the general
keyword `data science education.' We conducted a systematic literature
review \citep{evans2001systematic, liberati2009prisma} using criteria
that we detail in Section~\ref{sec-methodology}. In
Section~\ref{sec-results} we share our findings and then we discuss
their implications in Section~\ref{sec-discussion}.

\section{Data Collection and Analysis}\label{sec-methodology}

The target population of this literature review was publications on data
science education that directly address the undergraduate level. Due to
variations in terminology, keywords, and language used by
teacher-scholars across different data science fields, identifying the
entire target population was not feasible. Since there is no consensus
on what data science is, and consequently what data science education
is, we did not evaluate whether publications met a certain definition of
data science. Rather, we considered publications that self-proclaim to
focus on data science education. Therefore, we opted to identify the
accessible population for this study as publications that included
``data science education'' in quotes in the title, abstract, or
keywords.

\begin{figure}

\centering{

\pandocbounded{\includegraphics[keepaspectratio]{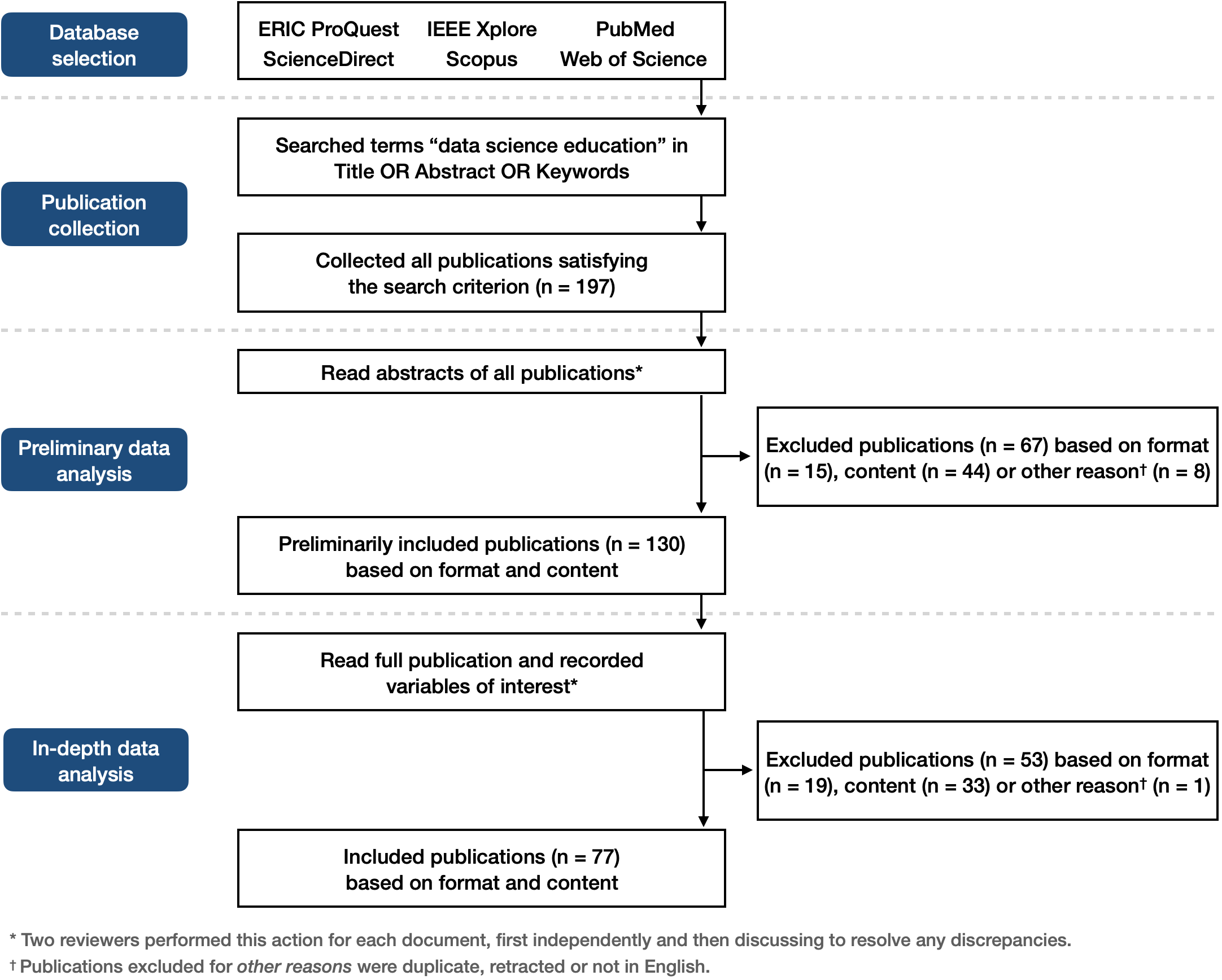}}

}

\caption{\label{fig-flowchart}Flowchart of data collection and analysis.
Publications were collected in December 2022.}

\end{figure}%

Figure~\ref{fig-flowchart} summarizes the stages of our data collection
and analysis processes which led to the sample of this study. As shown
in the diagram, we extracted data from six databases that potentially
include publications on data science education. These databases were:
(1) ERIC ProQuest, (2) IEEE Xplore, (3) PubMed, (4) ScienceDirect, (5)
Scopus, and (6) Web of Science. These databases were selected because
they cover a large number of publications and they include multiple
disciplines that are commonly linked to data science education. For
instance, ERIC ProQuest is known to mainly include education-focused
publications, IEEE Xplore focuses on engineering and Web of Science is
multidisciplinary-focused.

From the selected databases, we collected publications including the
term ``data science education'' (in quotes) in at least one of the
following fields: title, abstract, keywords. We acknowledge that there
are many other combinations of search terms that could result in
publications related to data science education, including terms such as
``data science courses'', ``data science pedagogy'', ``data science
curriculum''. One can generate numerous similar terms, with or without
quotes. We specifically used the term ``data science education'' for its
broadness and to set the scope of the research. We did not use ``data
science'' and ``education'' separately to avoid publications in
education that employ data science methods for their analyses.

Across the six databases, we found a total of 197 publications that met
our search criteria. We extracted some variables including, but not
limited to: author names; publication title; publication venue (e.g.,
journal title or conference title). We conducted the initial database
search in December 2022, resulting in a pool of publications that were
either published by that date or available online by that date but
officially published in 2023.

The data analysis was conducted in two stages: 1) preliminary data
analysis and 2) in-depth data analysis. Each publication was randomly
assigned to two authors of the present manuscript. At both data-analysis
stages, the two assigned reviewers first examined each publication
independently and then discussed discrepancies between their analysis
decisions, to reach a consensus. In cases where conflicts persisted, the
entire group of five authors deliberated on the final decision.

During preliminary data analysis, we manually opened and read the
abstracts of all publications. At this stage, we sought to exclude
publications that did not meet our format and content criteria based on
their abstracts. Specifically, we aimed to include
journal/conference/magazine articles and book chapters (\emph{format})
that focused on undergraduate data science education (\emph{content})
and were written in English. During in-depth analysis we examined the
full publications and, upon confirming that they met our inclusion
criteria, we recorded variables of potential interest. The number of
publications that were excluded due to format, content or other reason,
at each stage of our analysis, is shown in Figure~\ref{fig-flowchart}.

Across both stages, we excluded 34 publications due to formatting (which
included posters, panels, letters to journal editors and meeting
highlights). Of the remaining 163 publications, 77 were excluded due to
their content: 12 of them were not about data science education
(including e.g.~publications focused on data science methodology) and 65
of them focused exclusively on a different level of education than
undergraduate. Among the latter excluded group, 18 publications focused
on graduate level, 29 focused on K-12, middle school or high school and
18 publications focused on data science education for adults in
non-academic programs (including, e.g.~practitioners, citizen science
and instructors). Publications that focused on both undergraduate and
non-undergraduate levels were included. Of the remaining 86
publications, other 9 were excluded due to being: not written in
English; duplicated in our dataset; retracted by their authors.

After excluding publications for the reasons detailed above, we were
left with 77 publications, which we analyzed in-depth. In addition to
the variables extracted from databases (e.g.~title of the publication,
author names, etc.) for each publication we collected data on:

\begin{itemize}
\tightlist
\item
  affiliation country of researchers
\end{itemize}

\begin{itemize}
\item
  open access status (i.e.~whether the full publication is accessible
  for free from a Google Scholar search)
\item
  year when the publication was first published online
\item
  document type (conference article, journal article, magazine article
  or book chapter)
\item
  whether there were explicit research questions stated in the
  publication
\item
  whether there was any reporting of data collection in the publication
  and, if data were collected, the type of data (quantitative,
  qualitative or mixed)
\item
  publication focus, that is a categorization of the subject matter of
  the publication (for example ``pedagogical approach'', ``class
  activity'' or ``review of current state of data science education'')
\item
  discipline of the publishing source, determined by examining the call
  for contributions of the journal or conference, or the description of
  the book or magazine (``broad'' when the publishing source called for
  contributions across all data science fields, otherwise a specific
  sub-field of data science, e.g.~``computer science'' or
  ``statistics'')
\item
  the discipline of the target audience, as expressed by the authors in
  the publication (``broad'' when the publication target were all data
  science educators, otherwise a specific sub-field of data science,
  e.g.~``computer science'' or ``statistics'')
\end{itemize}

\section{Results}\label{sec-results}

\begin{figure}

\centering{

\pandocbounded{\includegraphics[keepaspectratio]{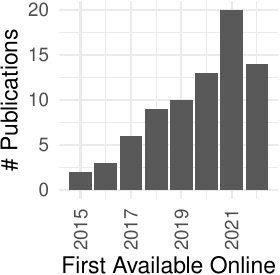}}

}

\caption{\label{fig-first_published}Undergraduate data science education
publications (\(n\) = 77) by year they were first made available online.
Note that initial data retrieval was in December 2022.}

\end{figure}%

\textbf{Publication years.} Figure~\ref{fig-first_published} shows the
distribution of the years when publications in our review were made
available online. Some of these publications were available during our
database search in December 2022, but would ultimately be published in a
2023 journal edition \citep[e.g.,][]{pieterman-bos2023, schmitt2023}.
The body of literature that we identified is very recent, with the
oldest paper available in 2015 and the volume of work increasing
steadily in the following years.

\textbf{Open access status and document types.} A majority of work on
undergraduate data science education published over these 8 years has
been freely available to the public: the breakdown of articles' open
access status is shown in Table~\ref{tbl-open_access} along with the
document type. The vast majority of journal articles were open access,
while conference publications showed a more even split. We reviewed only
four publications which were neither journal nor conference articles, of
which two were magazine articles \citep{hazzan2021, bonnell2022} and two
were book chapters \citep{manzella2022can, RYAN201645}.

\begin{longtable}[t]{llll}

\caption{\label{tbl-open_access}Document type stratified by open
access.}

\tabularnewline

\toprule
Document Type & Open Access & Not Open Access & Total\\
\midrule
\cellcolor{gray!10}{Conference Article} & \cellcolor{gray!10}{24} & \cellcolor{gray!10}{19} & \cellcolor{gray!10}{43 (56\%)}\\
Journal Article & 26 & 4 & 30 (39\%)\\
\cellcolor{gray!10}{Book Chapter} & \cellcolor{gray!10}{0} & \cellcolor{gray!10}{2} & \cellcolor{gray!10}{2 (3\%)}\\
Magazine Article & 1 & 1 & 2 (3\%)\\
\cellcolor{gray!10}{Total} & \cellcolor{gray!10}{51 (66\%)} & \cellcolor{gray!10}{26 (34\%)} & \cellcolor{gray!10}{77}\\
\bottomrule

\end{longtable}

\textbf{Affiliation countries.} We also investigated the geographic
breakdown of institutions that are contributing to literature on
undergraduate data science education. A majority of the publications
analyzed included at least 1 author associated with an American
institution (45 of 77), with European and Asian institutions providing
the bulk of the remaining analyzed literature.
Table~\ref{tbl-country-affiliation} breaks down the analyzed
publications by the institutional affiliations of their authors. The
higher number of scholars' affiliations being in the United States may
be at least partly attributed to the large size of the US relative to
many other countries, as well as selection bias due to our search being
restricted to English-written publications and to databases where US
scholars might be more represented.

\begin{longtable}[t]{lrlr}

\caption{\label{tbl-country-affiliation}For each country below, we
report the number of publications which included at least one affiliated
author. Affiliation country of authors was determined based on their
home institutions as reported in the publication, and does not represent
authors' nationalities.}

\tabularnewline

\toprule
Country & \# Publications & Country & \# Publications\\
\midrule
\cellcolor{gray!10}{United States of America} & \cellcolor{gray!10}{45} & \cellcolor{gray!10}{Australia} & \cellcolor{gray!10}{2}\\
Netherlands & 8 & Egypt & 2\\
\cellcolor{gray!10}{Canada} & \cellcolor{gray!10}{5} & \cellcolor{gray!10}{Germany} & \cellcolor{gray!10}{2}\\
Norway & 5 & Greece & 1\\
\cellcolor{gray!10}{United Kingdom} & \cellcolor{gray!10}{5} & \cellcolor{gray!10}{Hong Kong S.A.R.} & \cellcolor{gray!10}{1}\\
\addlinespace
Spain & 4 & Ireland & 1\\
\cellcolor{gray!10}{China} & \cellcolor{gray!10}{3} & \cellcolor{gray!10}{New Zealand} & \cellcolor{gray!10}{1}\\
India & 3 & Portugal & 1\\
\cellcolor{gray!10}{Israel} & \cellcolor{gray!10}{3} & \cellcolor{gray!10}{Romania} & \cellcolor{gray!10}{1}\\
Italy & 3 & Slovakia & 1\\
\addlinespace
\cellcolor{gray!10}{Japan} & \cellcolor{gray!10}{3} & \cellcolor{gray!10}{Thailand} & \cellcolor{gray!10}{1}\\
Switzerland & 3 & United Arab Emirates & 1\\
\bottomrule

\end{longtable}

\begin{figure}

\centering{

\pandocbounded{\includegraphics[keepaspectratio]{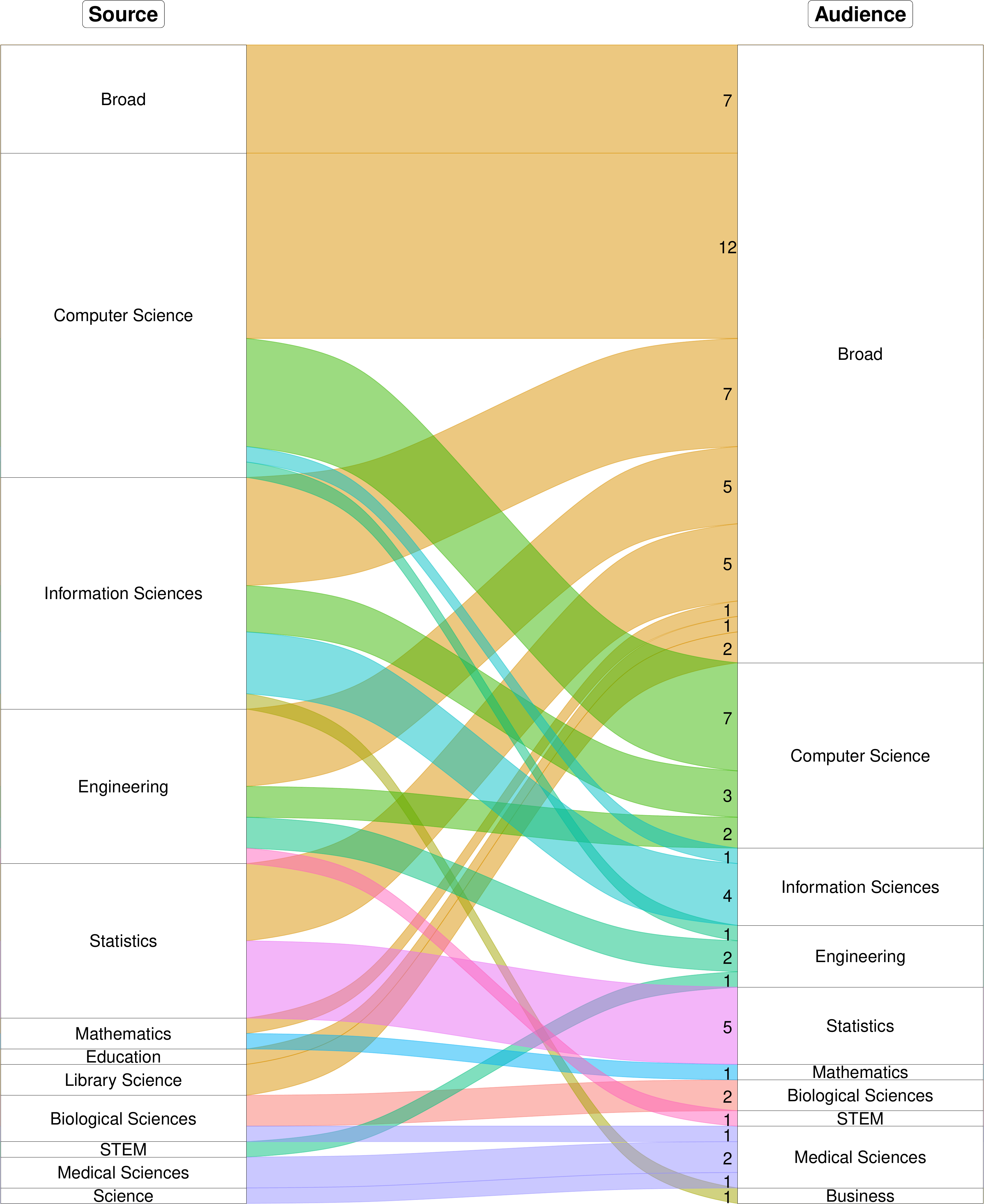}}

}

\caption{\label{fig-discipline}Composition of the discipline of targeted
audiences of publications on data science education (right) by the
discipline of the publishing source (left). A total of 2 publications
were not included in the plot, one because it stated to target both
statistics and computer science, the other one because it was not
possible to determine its audience discipline.}

\end{figure}%

\textbf{Source and audience disciplines.} Figure~\ref{fig-discipline}
shows the breakdown of articles by discipline. We report both the
discipline of the publishing venue (such as the journal's or
conference's discipline), and the audience that the publication stated
(or implied) to target. About half of the articles (40/77) were written
for the broad community of data science educators. The remaining
articles were written for educators in particular disciplines, most
commonly computer science (12 publications), statistics (5
publications), and information sciences (5 publications). We found that
publication venues in all of these disciplines, as well as engineering,
contributed articles aimed at their specific audience as well as
articles aimed at the broad community of data science educators.

\textbf{Research Question and Data Collection.} Most publications (69 of
77 reviewed) either posed research questions \emph{and} collected data
or did neither, although there were exceptions. Four publications had
research questions but no data collected
\citep{vance2022, pieterman-bos2023, hagen2020, robeva2020}. On the
other hand, four publications collected data but did not pose a specific
research question or study goal
\citep{hicks2018, rao2018, liu2020, cuadrado-gallego2021}. Of the 40
publications that included collected data, 6, 9, and 25 publications had
qualitative, quantitative, and mixed data respectively.

\begin{figure}

\centering{

\pandocbounded{\includegraphics[keepaspectratio]{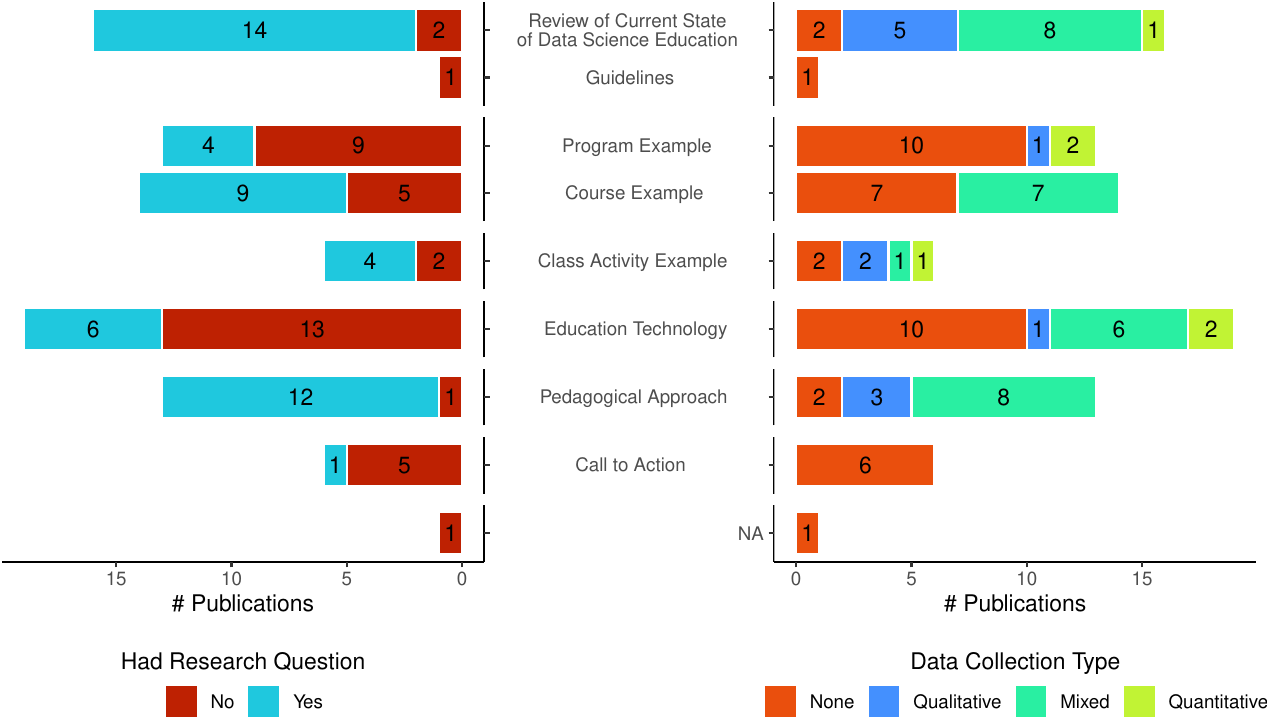}}

}

\caption{\label{fig-publication-focus}Number of publications with
communicated research questions and with collected data stratified by
publication focus.}

\end{figure}%

\textbf{Publication focus.} We also classified publications based on
their focus. We found publications that reviewed the current state of
data science education and provided guidelines. We also encountered
publications that provided examples of programs, courses, class
activities, and extracurricular activities. Some publications
specifically focused on educational technology or discussed a
pedagogical approach. Some publications included more than one focus: in
such cases, we noted down all the categories that the publication falls
into, with the exception of \emph{call to action}. Almost all
publications we read were calling the scientific community to action and
this is unsurprising for scientific publications. For instance, an
educational technology paper might call to action to use a specific tool
and a pedagogy paper might call to action to use a specific pedagogical
approach. However, some publications solely made a call to action,
without content falling into other publication focuses (e.g., education
technology or course example). For these publications, we reserved the
publication focus ``call to action'', that is, papers that we labeled in
the ``call to action'' category did not fall in any other category for
publication focus.

For each category of publication focus,
Figure~\ref{fig-publication-focus} shows how many publications featured
that publication focus, had a research question and included
qualitative, quantitative or mixed type of data. Note that, in this
figure, publication focuses are arranged and grouped to match the order
in which we are going to illustrate them below, as we highlight
contributions from each publication focus.

Similar to what we are doing in our study, many scholars wanted to
understand the \textbf{current state of data science education}. Many of
the studies conducted wanted to understand and evaluate data science
programs
\citep{wiktorski2016, oliver2021, song2016, shao2021, li2021, raj2019}
or curricula \citep{schmitt2023}. \citet{davis2020} specifically looked
at the current state of ethics in undergraduate data science education
and \citet{ceccucci2015} considered data science education from a
scientific literacy perspective. Among these reviews, there were also
comparative studies: for instance, \citet{bilehassan2020}'s comparison
of informatics and data science programs. Some studies also compared
differences of data science approaches at country or regional levels
such as in Japan \citep{takemura2018}, or in Middle Eastern countries
\citep{zakaria2023}. In understanding the current state of data science
education, scholars also wanted to understand the data science practice.
For example, \citet{belloum2019} developed semi-automated methods to
determine the competencies needed in the job-market and
\citet{kross2019} focused on understanding the skills that data science
practitioners who teach data science in various settings pass onto their
students. In addition to reviewing data science education, scholars also
provided \textbf{guidelines} for data science degrees \citep{blair2021}.

We found multiple data science \textbf{program examples}, including but
not limited to general data science programs
\citep[e.g.,][]{demchenko2017, demchenko2019, kakeshita2022}. Some
publications were about data science education in specific programs such
as computer science education \citep{bilehassan2020}, microbiology
\citep{dill-mcfarland2021}, information schools
\citep{song2017, hagen2020} and business \citep{miah2020}. Less
traditional programs were also featured, such as the Data Mine which
takes on data science education in a residential community of students
\citep{betz2020}.

The literature we read also included many \textbf{course examples} with
some clever ways of including data science concepts in different
courses. For instance, \citet{fisler2022} suggested to include data
science and data structures in an introductory computing course,
\citet{rao2019} teach data science through the use of education data in
their engineering course on modern technologies, and \citet{haynes2019}
teach data science in a general education IT course. Some institutions
developed data science courses for specific disciplines including
information schools \citep{hagen2020} and medicine \citep{doudesis2022}.
Some scholars also described courses that merge data science with
philosophy of science \citep{pieterman-bos2023} or with humanities
perspectives \citep{vance2022}. Last but not least, data science
educators also tried to provide real-life data science experiences
through work-integrated learning \citep{bilgin2022}, capstone projects
\citep{allen2021} and entirely case-study based courses
\citep{hicks2018}.

Similar to program examples and course examples provided in the
literature, some publications focused on activity examples. Among
\textbf{class activity examples}, \citet{yamamoto2021} developed a
programming exercise to bring higher-order tensor content to
undergraduate level by using a 3-D puzzle. Another example is by
\citet{banadaki2022} who developed different activities that include
applications of data science in Mechanical Engineering, Biomedical
science, Quantum Physics, and Cybersecurity. In data science education,
there is also room for learning outside the classroom as an
\textbf{extra-curricular activity}. This included data hackathons
\citep{anslow2016}.

In terms of \textbf{education technology}, as much of data science
education tackles with issues such as computing power, storage of large
datasets, and sometimes automation, some of these publications focused
on cloud-based data science platforms for teaching purposes
\citep[e.g.,][]{demchenko2017}. Many others focused on tools such as
online platforms, learning environments, and apps that support learning
data science
\citep[e.g.,][]{hoyt2018, bornschlegl2016, liu2020, nair2020}.

Studies that focused on \textbf{pedagogical approaches} were also
frequently encountered. These studies focused on various subtopics
including team-based learning \citep[e.g.,][]{vance2022}, project-based
learning \citep[e.g.,][]{mike2020} and large-scale product development
which has both project- and team-based learning aspects
\citep[e.g.,][]{bhavya2020}. Scholars also studied social topics such as
ethics and equity in the data science classroom \citep{alexander2022}
and student self-regulation \citep{zhang2020}.

Almost all publications we read were calling the scientific community to
action. For instance, through a systematic review, \citet{davis2020}
\textbf{called the community to action} to include ethics in
undergraduate data science education. We also encountered publications
with an important message about data science education but without a
review of programs, courses, activities etc. For instance,
\citet{robeva2020} argued for the inclusion of data science in
quantitative biology education. \citet{engel2017} drew attention to the
importance of statistical literacy in data science education. We have
also seen a to-do list (i.e., action list) for the community including
starting a multidisciplinary data science education journal
\citep{hazzan2021}, building a consensus on data science education and
curricula \citep{dinov2019}, developing and deploying degree programs
for university students, having basic data science training for
university students, and training instructors to teach data science
\citep{bonnell2022}.

\section{Discussion}\label{sec-discussion}

Turning our attention to the knowledge that was gathered and summarized
in our study, this section examines the strengths and gaps in the
literature authored by researchers who self-proclaim to focus on
undergraduate data science education. Building upon the findings
detailed in Section~\ref{sec-results}, we also extend comprehensive
recommendations to both policymakers and the data science education
community, aiming to enhance capacity building in undergraduate data
science education studies.

Before proceeding, we believe that it is essential to address the
limitations (1) inherent in the field of data science and (2) related to
the scope of this study. Acknowledging these constraints provides
crucial context and helps avoid overgeneralization, by enabling readers
to more effectively contextualize the strengths, gaps, and
recommendations presented in this section.

\subsection{Inherent Limitations in the Field of Data Science
(Education)}\label{inherent-limitations-in-the-field-of-data-science-education}

Data science is an interdisciplinary \citep{cao2017} and relatively
young field. While its emergent nature brings numerous opportunities, it
also introduces inherent limitations that can influence the accumulation
of knowledge in this field and, consequently, the findings of this
systematic literature review.

Perhaps the most critical limitation is the lack of consensus on the
definition of data science \citep{hazzan2023}. Although we acknowledge
that the absence of a standard definition is not unique to the field of
data science, this lack of agreement at such an early stage, especially
in a field that is so interdisciplinary, may result in a deeper issue:
the challenge of generating an \emph{identifiable} cumulative body of
knowledge in this field.

Indeed, a corollary of this definitional ambiguity is the difficulty in
labeling research as data science education. Some studies may be
considered data science education by certain scholars, but lack explicit
labeling as such, while others that we included might not be categorized
as data science education under different interpretations. Therefore,
capturing all undergraduate data science education research papers is
extremely challenging at this stage.

Arguably, the lack of clear identifiability and consistent self-labeling
creates at least two additional challenges. First, it restricts
practitioners' access to findings that could inform and enhance their
data science teaching practices. Second, it may hinder researchers from
identifying recurring patterns and comprehensively understanding the
current state of data science education research.

In this study, we deliberately refrained from defining data science and
data science education. Instead, we used ``data science education'' as a
keyword to capture publications that self-proclaim as undergraduate data
science education research, across all disciplines. We opted for this
approach because we did not want to take the role of gate-keepers by
assessing whether publications adhered to any specific definition of
data science. However, this choice may lead to an \emph{incomplete
picture} of the undergraduate ``data science education'' literature, as
we discuss in the subsequent section.

\subsection{Limitations of the Scope of This
Study}\label{limitations-of-the-scope-of-this-study}

We acknowledge that, in this study, we did not reach the entire target
population of research publications on undergraduate data science
education. In addition to the inherent limitations of the field outlined
above, methodological limitations may also contribute to this incomplete
picture of undergraduate data science education literature.

One key methodological limitation is the use of a single term, ``data
science education'', for our search. While the choice of this term was
aimed at capturing a broad range of self-proclaimed undergraduate data
science education research, it may have restricted the results. For
instance, an article that includes terms such as ``data science class'',
``data science activity'', or ``data science curriculum'' but does not
use the keyword ``data science education'' would be excluded from this
study \citep[e.g.,][]{baumer2022integrating}. Definitional ambiguities
and self-labeling challenges in the field may exacerbate this issue.

Our review may have missed relevant publications due to several other
factors, including publications written in languages other than English
and publications that are not included in the databases that we searched
\citep[e.g.,][]{finzer2013data}.

Despite these challenges, we believe our study offers valuable insights
into the strengths and knowledge gaps within self-proclaimed
undergraduate data science education research. It provides a realistic
portrayal of what data science educators might encounter when searching
for literature using the general keyword ``data science education.''
Furthermore, it offers insights that can benefit the broader data
science education community, including those whose work does not
explicitly label itself as data science education.

\subsection{Strengths in Undergraduate Data Science
Education}\label{strengths-in-undergraduate-data-science-education}

The majority of published studies on undergraduate data science
education are open access, marking a substantial strength in the field.
The freely available information and scholars' insights regarding the
status of undergraduate data science education not only contribute to
overcoming barriers, but also facilitate the dissemination and
application of knowledge. Over the eight years included in our analysis,
a higher percentage of conference articles (56\%) have been published
compared to journal articles (39\%). Education technology is the
publication focus that is studied most by scholars. Among educational
technologies to support learning data science, learning environments
\citep[e.g.,][]{bornschlegl2016, hoyt2018} are one of the popular ones.
Ethics is also one of the recurring themes among the studies that we
reviewed \citep[e.g.,][]{davis2020, shapiro2020, alexander2022}.

As shown in Figure~\ref{fig-discipline}, the majority of self-proclaimed
undergraduate data science publications are directed to the broad
audience of data science educators generally. Specifically, 40 out of 77
publications target the undergraduate data science education community
as a whole, rather than appealing to a specific discipline in data
science (e.g., computer science) or a subset of disciplines in data
science (e.g., STEM). This inclusivity can be seen as a potential
strength, as the insights from these broader publications provide
valuable perspectives that scholars and researchers can adapt and apply
across diverse contexts in data science to enhance teaching practices,
develop curricula, and foster a more comprehensive understanding of data
science education.

Studies including multiple examples of data science programs and courses
across various fields add another layer of strength to the undergraduate
data science education literature. In addition to overall data science
programs \citep[e.g.,][]{demchenko2019, kakeshita2022}, we have also
seen data science education practices in different programs such as
computer science education \citep{bilehassan2020}, microbiology
\citep{dill-mcfarland2021}, and business \citep{miah2020}.

The course examples showcase a diverse array of strategies for
incorporating data science concepts, ranging from introductory computing
\citep{fisler2022}, modern technologies for computer science and
engineering students \citep{rao2019} and general education IT
(Information Technology) \citep{haynes2019}, to medicine
\citep{doudesis2022} and introduction to psychological statistics
\citep{tucker2023}. This broad spectrum in both program and course
examples serves as compelling evidence, illustrating that the intrinsic
interdisciplinary nature of data science attracts the attention of
scholars from diverse fields.

\subsection{Knowledge Gaps in Undergraduate Data Science
Education}\label{knowledge-gaps-in-undergraduate-data-science-education}

While the strengths of the current data science education literature, as
outlined above, are evident, our study also reveals some knowledge gaps.
Knowledge gaps are areas or topics derived from the synthesis of an
existing body of literature \citep{cooper1998syn}. Understanding these
gaps is crucial because it adds a more structured and evidence-supported
layer to our knowledge. In this section, we discuss some knowledge gaps
in self-proclaimed undergraduate data science education.

\textbf{Knowledge gap 1: Certain disciplines in data science are less
visible in the current body of literature.}

Despite the lack of consensus on defining what exactly data science is,
there is agreement that data science encompasses various disciplines,
including statistics, mathematics, computer science, and other relevant
domains, as defined by many interpretations
\citep[e.g.,][]{cao2017, hazzan2023}. However, these disciplines are not
equally represented among the self-proclaimed data science education
publications that we examined.

In terms of source discipline (i.e., the main discipline of a journal,
conference, book chapter etc.) we have seen many publications published
in venues related to computer science, information science, engineering,
and statistics. The difference in quantity of publications published in
sources related to computer science and statistics is worth remembering
from Figure~\ref{fig-discipline}. This result partly aligns with the
study of \citet{wiktorski2017}, who reported that mathematics and
statistics departments are not at the forefront of data science. This
result can also potentially be explained by the fact that different
disciplines have different publication rates in general. For instance,
in the \emph{Science and Technology Indicators} report published by NSF
it is stated that 2.3\% of the articles published in 2016 is in the
field of mathematics and 8.3\% are in computer sciences
\citeyearpar{nsfindicators}.

We have seen even fewer publications in venues related to domain
sciences. These were mainly related to education, biological sciences,
library science, and medical sciences. The findings may indicate a need
to have journals and conferences in other domain sciences (e.g.,
astronomy, economics, psychology) that provide more opportunities for
disseminating works of data science education researchers to a broader
data science community. It is also possible that such journals and
conferences for domain sciences might exist, but the keyword ``data
science education'' is not used in these venues. For instance,
economists may continue to use the term ``econometrics education''
rather than ``data science education''.

Among source disciplines, we have also identified publication venues for
the broader data science community. This, however, mainly consisted of
conferences. The two journals that were identified as broad were
\emph{International Journal of Data Science and Analytics} and
\emph{Foundations of Data Science}. Both of these journals focus on data
science while allowing for education-related publications, but neither
of these journals focuses specifically on data science education. These
findings reiterate the importance of the call made by Hazzan and Mike
\citeyearpar{hazzan2021} on having an interdisciplinary journal on data
science education.

In terms of audience disciplines, scholars write for the broader data
science education community as well as specific disciplines.
Representation of both broad and specific disciplines is important and
can enrich the data science education research. There are fewer
publications written for the audiences in domain sciences. This might
indicate further need for data science education research targeting
these audiences, or this might be due to under-utilization of the
keyword ``data science education'' in these disciplines.

Statistics is underrepresented both as source and audience discipline in
the body of literature that we reviewed. Arguably, statisticians are
contributing to data science education research more than this, but
their work was not prominently captured in this study. Several
explanations may account for this finding. Statisticians might avoid
using the term ``data science'' in their work, perceiving their research
as strictly related to statistics and targeting their primary audience
within that domain. Another possibility is that they consider data
science as a subset or natural extension of statistics, and thus do not
see the need to explicitly label their work as data science education
research. Lastly, they might be using some other keywords in their
title, abstract, and keywords instead of ``data science education'',
making it less visible to the broader data science education community.

This limited visibility of statisticians in the current literature
underscores the insufficiency of the existing body of data science
education research. Most importantly, it highlights a critical gap: the
lack of visible perspectives from statisticians in this identified body
of literature. As statistics is one of the foundational disciplines in
data science, its under-representation not only diminishes the diversity
of insights but may also impede the development of comprehensive,
interdisciplinary approaches to data science education.

Although statistics is underrepresented both as source and audience
discipline in comparison to computer science, it is worth noting that
both computer science and statistics sources write for broad data
science education audiences as well as their corresponding specific
discipline audiences. For foundational sciences of data science, such as
statistics and computer science, writing for the broader data science
community as well as the specific discipline is important and can
continue to enrich data science education research in the years to come.

\textbf{Knowledge gap 2: Within the identified body of data science
education literature, there is lack of empirical data and identifiable
research questions.}

Research questions and data are two common elements in empirical
research. Data for empirical research include both qualitative and
quantitative approaches where the researchers collect `observable
information about or direct experience of the world'. And perhaps most
importantly, empirical data are not just stored as in numbers but also
words and categories \citep{punch2014introduction}.

One of the important functions of systematic literature reviews is to
gain a deeper understanding and inform possible further research avenues
that can be conducted in the field
\citep{evans2001systematic, liberati2009prisma}. To facilitate this, we
categorized the studies based on their publication focus, the existence
of research questions, and the types of data collection. As stated
earlier, 37 studies out of 77 did not collect data. Given the scopes of
publication focus such as calls to action, educational technology, and
program examples coupled with the emergent nature of the field, the lack
of empirical data is not a surprising finding. However, it also suggests
that undergraduate data science education researchers have not yet begun
to systematically collect empirical data to assess, for example, the
effectiveness of educational technologies, programs, learning outcomes
and/or other pedagogical approaches.

A lack of empirical data could impede the development of systematic
literature reviews or meta-analyses on a specific publication focus
(e.g., educational technology), which are essential for identifying
trends, studied variables, and recurring patterns in undergraduate data
science education through qualitative and/or quantitative approaches. It
is essential to clarify that our emphasis on lack of empirical data is
not a promotion of empiricism over all other `ways of knowing'. We
acknowledge and appreciate alternative forms of knowledge, such as
expert opinions \citep{fraenkel2012design}, for their valuable
contributions to the data science education community's know-how. These
forms of knowledge are important catalysts in guiding researchers
towards areas that require systematic data collection. What we are
highlighting is the disproportionately high percentage of studies
lacking empirical data and identifiable research questions, which
complicates the literature's potential for gaining a deeper
understanding and identifying recurring patterns.

\textbf{Knowledge gap 3: Reproducibility is one of the potential
challenges in undergraduate data science education research.}

The corollary of a lack of empirical data and identifiable research
questions may introduce another challenge: the reproducibility of
certain studies. We speculate that the absence of critical information
about research designs, such as the lack of research questions and
non-collection of data, may contribute to the reduced reproducibility of
available studies. This makes it challenging to replicate or modify
research, impeding the identification of recurring patterns.

Lack of reproducibility is not unique to data science education
research. Its importance and lack thereof have been discussed in many
disciplines including physics \citep{junk2020}, economics \citep{chang},
and psychology \citep{opensciencecollaboration2015}. In recent years,
this has even been referred to as the \emph{reproducibility crisis}.

Potential explanations for the lack of reproducibility in data science
education research might be similar to reasons seen in the broader
science community. Namely, word limitations can result in lack of
detailed information on research design and data collection in
publications \citep{bausell2021}.

Another explanation might simply be the ``publish or perish'' culture in
academic settings. There are even academics who publish a paper every
five days \citep{ioannidis2018}. Even if not at this rate, many
academics might feel under pressure to get publications out, without
having much time to focus on the reproducibility of their work.

Reproducibility is an important skill for teacher-scholars of data
science both in their teaching and research
\citep{dogucu2022, dogucu2024}. Considering that much of the published
research in the literature are written by those who also teach data
science, closing the reproducibility gap both in research and teaching
is extremely important. One potential reason for this gap may also be
the minimal training that most instructors receive in reproducibility
\citep{horton2022}.

\subsection{Recommendations}\label{recommendations}

Considering the findings and our arguments, we present three
recommendations for the future of undergraduate data science education
studies: one for policymakers and funding agencies, and two for
institutions and scholars whose research focus includes data science
education.

\textbf{Recommendation 1: Cherish data science as an interdisciplinary
field.}

Despite the existence of other key fields offering data science courses
at the undergraduate level, there is a noticeable gap in studies
reflecting their perspectives in data science education. To address this
gap, it is imperative for undergraduate data science education studies
to incorporate the viewpoints of scholars from foundational disciplines
such as statistics, mathematics, and other application domains
intersecting with data science education practices. We need to encourage
scholars in all data science fields to maintain their visibility and
contribute more to publications. We posit that future endeavors in this
direction will significantly enhance our understanding of the strengths
and needs in undergraduate data science education.

As statistics community, we must take an active role in expanding the
data science education research and other opportunities. For instance,
despite having Guidelines for Assessment and Instruction in Statistics
Education (GAISE) \citeyearpar{GAISE2016} and Curriculum Guidelines for
Undergraduate Programs in Statistical Science (CGUPSS)
\citeyearpar{asacurricula}, ASA has not yet written any guidelines
specific to data science education. However, ASA Board of Directors
endorsed the Park City Math Institute report \citep{de2017curriculum}
and have provided input on the criteria for the data science program
ABET accreditation \citep{liu2022}. At the time of writing of this
manuscript, ASA is working to update GAISE. Further efforts by ASA
similar to GAISE and CGUPSS or newer versions of these documents can
also help distinguish statistics and data science programs and courses.

\textbf{Recommendation 2: Adopt a consistent set of keywords/terminology
to ensure data science education literature is easily identifiable.}
While there is no doubt that all data science fields are contributing to
data science education research, some of their work was not prominently
captured in this study. Speculatively, it appears that the keyword
``data science education'' is more commonly embraced by certain fields
compared to others. As a result, a researcher broadly interested in data
science education may not see the interdisciplinary diversity of
insights within the field, if there is a lack of consistent use of
keywords and terminology.

We believe that using both broad, cross-disciplinary terms and
specialized terminology unique to each field/domain is essential for
facilitating communication within individual disciplines/domains and
across the wider data science community. Adopting a consistent set of
keywords does not require a shared definition of what is and is not data
science. If a set of keywords is consistently used by data science
education researchers, it may facilitate the accumulation of collective
knowledge and help identify what works or does not in the field.
Therefore, it is important to standardize terminology to enhance the
accessibility and comprehensiveness of the data science education
literature.

\textbf{Recommendation 3: Prioritize investments in empirical studies.}
Not having sufficient empirical data is one of the knowledge gaps in
undergraduate data science education research. Accumulating empirical
data is essential to be able to gain a sound understanding of data
science education studies at a large scale. Hence, we recommend that
policymakers and funding agencies prioritize investments in
undergraduate data science studies dedicated to systematic data
collection. Additionally, directing investments toward empirical studies
in application domains of data science that were underrepresented in our
study, such as astronomy, psychology, and economics could also help
provide a more complete picture of data science education in the long
run.

These strategic approaches will enable a comprehensive assessment of
foundational disciplines and various application domains of data
science, including the effectiveness of educational technologies,
program impact, learning outcomes, and other pedagogical approaches,
ultimately contributing to a more informed and robust understanding of
data science education.

\subsection{Closing Remarks}\label{closing-remarks}

In summary, the results of this study show that data science education
is an emerging field with much more room for growth. Scientific studies
are an integral part of reviewing existing practices as well as of
improving higher education institutions' data science practices.
Therefore, we should diversify our research efforts by investing in more
empirical studies and fostering scholars from key fields in data
science, especially in statistics and domain sciences.

Further research studies may improve or try to replicate the findings of
this study in multiple ways by utilizing different keywords, databases,
and languages. An even larger endeavor can be taken if investigators
want to work off on a definition of data science utilizing a systematic
literature review.

Lastly, we also believe that the data collected as part of this study
can help novice researchers in data science education find inspiration
and examples from the existing literature. We presented findings of the
study at the Electronic Conference on Teaching Statistics 2024 where we
had participants design their own research study by grouping them
according to their publication focus (e.g., educational technology) of
interest. Readers interested in pursuing research in data science
education may utilize our dataset to create a reading list for their
research agenda.

\section*{Data Availability
Statement}\label{data-availability-statement}
\addcontentsline{toc}{section}{Data Availability Statement}

The data on all the publications and the associated codebook for the
variables are publicly available in a GitHub repository at
https://github.com/mdogucu/comp-data-sci and an OSF project at
https://osf.io/b3u7y/.

  \bibliography{references.bib}

@online{nsf,
author = {{National Science Foundation}},
title = {{{NSF}'s 10 Big Ideas}},
year = {2017},
url={\url{https://www.nsf.gov/news/special_reports/big_ideas/}},
urldate={2023-09-11}
}

@online{abet,
author = {{ABET}},
title = {{Criteria for Accrediting Applied and Natural Science Programs, 2024-2025}},
year = {2024},
url={\url{https://www.abet.org/accreditation/accreditation-criteria/criteria-for-accrediting-applied-and-natural-science-programs-2024-2025/}},
urldate={2024-01-23}
}

@article{liberati2009prisma,
  title={{The PRISMA Statement for Reporting Systematic Reviews and Meta-Analyses of Studies that Evaluate Health Care Interventions: Explanation and Elaboration}},
  author={Liberati, Alessandro and Altman, Douglas G and Tetzlaff, Jennifer and Mulrow, Cynthia and G{\o}tzsche, Peter C and Ioannidis, John PA and Clarke, Mike and Devereaux, Philip J and Kleijnen, Jos and Moher, David},
  journal={Annals of Internal Medicine},
  volume={151},
  number={4},
  pages={W--65},
  year={2009},
  publisher={American College of Physicians}
}

@article{evans2001systematic,
  title={{Systematic Reviews of Educational Research: Does the Medical Model Fit?}},
  author={Evans, Jennifer and Benefield, Pauline},
  journal={British Educational Research Journal},
  volume={27},
  number={5},
  pages={527--541},
  year={2001},
  publisher={Wiley Online Library}
}

@article{asajour2020,
	title = {{ASA Journal Gets New Name, Mission}},
	author= {Witmer, Jeff},
	journal = {Amstat News},
	year = {2020},
	issue = {522},
	url = {\url{https://magazine.amstat.org/blog/2020/12/01/journal-gets-new-mission/}},
}

@article{de2017curriculum,
  title={{Curriculum Guidelines for Undergraduate Programs in Data Science}},
  author={De Veaux, Richard D and Agarwal, Mahesh and Averett, Maia and Baumer, Benjamin S and Bray, Andrew and Bressoud, Thomas C and Bryant, Lance and Cheng, Lei Z and Francis, Amanda and Gould, Robert and others},
  journal={Annual Review of Statistics and Its Application},
  volume={4},
  pages={15--30},
  year={2017},
  publisher={Annual Reviews}
}

@inproceedings{wiktorski2017,
  title={{Model Curricula for Data Science EDISON Data Science Framework}},
  author={Wiktorski, Tomasz and Demchenko, Yuri and Belloum, Adam},
  booktitle={2017 IEEE International Conference on Cloud Computing Technology and Science (CloudCom)},
  pages={369--374},
  year={2017},
  organization={IEEE}
}

@article{stern2021data,
  author = {Stern, Hal S. and Richardson, Debra J. and Papaefthymiou, Marios},
  title = {Data Science and Computing: The View From a Sister Campus},
	journal = {Harvard Data Science Review},
	number = {2},
	year = {2021},
	month = {jun 30},
	volume = {3}
}

@article{adhikari,
  title={{Interleaving Computational and Inferential Thinking in an Undergraduate Data Science Curriculum}},
  author={Adhikari, Ani and DeNero, J and Jordan, MI},
  journal={Harvard Data Science Review},
  volume={3},
  number={2},
  year={2021},
  doi={10.1162/99608f92.cb0fa8d2}
}

@article{baumer,
  title={{A Data Science Course for Undergraduates: Thinking with Data}},
  author={Baumer, Ben},
  journal={The American Statistician},
  volume={69},
  number={4},
  pages={334--342},
  year={2015},
  doi={10.1080/00031305.2015.1081105}
}

@article{glantz,
  title={{Students' Experience and Perspective of a Data Science Program in a Two-Year College}},
  author={Glantz, Mary and Johnson, Jennifer and Macy, Marilyn and Nunez, Juan and Saidi, Rachel and Velez Ramirez, Camilo},
  journal={Journal of Statistics and Data Science Education},
  volume={31},
  number={3},
  pages={248--257},
  year={2023},
  publisher={Taylor \& Francis},
  doi={10.1080/26939169.2023.2208185}
}

@incollection{manzella2022can,
  title={{How Can Ocean Science Observations Contribute to Humanity?}},
  author={Manzella, Giuseppe MR and Emery, William},
  booktitle={Ocean Science Data},
  pages={319--335},
  year={2022},
  publisher={Elsevier}
}

@incollection{RYAN201645,
title = {{From Self-Service to Self-Sufficiency}},
editor = {Lindy Ryan},
booktitle = {The Visual Imperative},
publisher = {Morgan Kaufmann},
address = {Boston},
pages = {45-60},
year = {2016},
isbn = {978-0-12-803844-4},
doi = {https://doi.org/10.1016/B978-0-12-803844-4.00003-0},
author = {Lindy Ryan}
}

@inproceedings{demchenko2017,
title = {{Customisable Data Science Educational Environment: From Competences Management and Curriculum Design to Virtual Labs On-Demand}},
	booktitle = {2017 {IEEE} International Conference on Cloud Computing Technology and Science (CloudCom)},
	author = {Demchenko, Yuri and Belloum, Adam and de Laat, Cees and Loomis, Charles and Wiktorski, Tomasz and Spekschoor, Erwin},
	year = {2017},
	month = {12},
	date = {2017-12},
	pages = {363--368},
	doi = {10.1109/CloudCom.2017.59}
}

@article{hoyt2018,
	title = {{An Overview of Two Open Interactive Computing Environments Useful for Data Science Education}},
	author = {Hoyt, Robert and Wangia-Anderson, Victoria},
	year = {2018},
	month = {10},
	date = {2018-10-01},
	journal = {JAMIA Open},
	pages = {159--165},
	volume = {1},
	number = {2},
	doi = {10.1093/jamiaopen/ooy040}
}

@inproceedings{rao2018,
  title = {{Milo: A Visual Programming Environment for Data Science Education
}},
	booktitle = {2018 IEEE Symposium on Visual Languages and Human-Centric Computing (VL/HCC)},
	author = {Rao, Arjun and Bihani, Ayush and Nair, Mydhili},
	year = {2018},
	month = {10},
	date = {2018-10},
	pages = {211--215},
	doi = {10.1109/VLHCC.2018.8506504}
}

@inproceedings{bornschlegl2016,
	title = {{IVIS4BigData: Qualitative Evaluation of an Information Visualization Reference Model Supporting Big Data Analysis in Virtual Research Environments}},
	booktitle = {Advanced Visual Interfaces. Supporting Big Data Applications: AVI 2016 Workshop},
	author = {Bornschlegl, Marco Xaver},
	editor = {Bornschlegl, Marco X. and Engel, Felix C. and Bond, Raymond and Hemmje, Matthias L.},
	year = {2016},
	date = {2016},
	pages = {127--142},
	series = {Lecture Notes in Computer Science},
	doi = {10.1007/978-3-319-50070-6_10},
	address = {Cham},
	langid = {en}
}

@inproceedings{shapiro2020,
	title = {Re-Shape: A Method to Teach Data Ethics for Data Science Education},
	author = {Shapiro, Ben Rydal and Meng, Amanda and {O'Donnell}, Cody and Lou, Charlotte and Zhao, Edwin and Dankwa, Bianca and Hostetler, Andrew},
	year = {2020},
	month = {04},
	date = {2020-04-23},
	pages = {1{\textendash}13},
	booktitle = {CHI '20},
	doi = {10.1145/3313831.3376251}
}

@inproceedings{liu2020,
  title = {How to Use Stock Data for Data Science Education: A Simulated Trading Platform in Classroom},
	booktitle = {2020 IEEE 2nd International Conference on Computer Science and Educational Informatization (CSEI)},
	author = {Liu, Yunkai and Wei, Xiangjing},
	year = {2020},
	month = {06},
	date = {2020-06},
	pages = {5--8},
	doi = {10.1109/CSEI50228.2020.9142534}
}

@inproceedings{nair2020,
	title = {{MetaData: A Tool to Supplement Data Science Education for the First Year Undergraduates}},
	author = {Nair, Rahul and Chugani, Mukesh N and Thangavel, Senthil Kumar},
	year = {2020},
	month = {05},
	pages = {153{\textendash}160},
	booktitle = {ICIET 2020},
	doi = {10.1145/3395245.3396409},
	address = {New York, NY, USA}
}

@inproceedings{mike2020,
title = {{Interdisciplinary Education - The Case of Biomedical Signal Processing
}},
	booktitle = {2020 IEEE Global Engineering Education Conference (EDUCON)},
	author = {Mike, Koby and Nemirovsky-Rotman, Shira and Hazzan, Orit},
	year = {2020},
	month = {04},
	date = {2020-04},
	pages = {339--343},
	doi = {10.1109/EDUCON45650.2020.9125200}
}

@article{alexander2022,
	title = {{Beyond Ethics: Considerations for Centering Equity-Minded Data Science}},
	author = {Alexander, Nathan and Eaton, Carrie Diaz and Shrout, Anelise and Tsinnajinnie, Belin and Tsosie, Krystal},
	year = {2022},
	month = {07},
	date = {2022-07-31},
	journal = {Journal of Humanistic Mathematics},
	pages = {254--300},
	volume = {12},
	number = {2},
	doi = {10.5642/jhummath.OCYS6929}
}

@article{vance2022,
	title = {{Integrating the Humanities into Data Science Education}},
	author = {Vance, Eric A and Glimp, David R and Pieplow, Nathan D and Garrity, Jane M and Melbourne, Brett A},
	year = {2022},
	date = {2022},
	journal = {Statistics Education Research Journal},
	pages = {9--9},
	volume = {21},
	number = {2}
}

@inproceedings{zhang2020,
	title = {{Self and Socially Shared Regulation of Learning in Data Science Education: A Case Study of {\textquotedblleft}Quantified Self{\textquotedblright} Project}},
	author = {Zhang, Jiangxiang and Wu, Bian},
	year = {2020},
	month = {06},
	date = {2020-06},
	booktitle={ICLS 2020 Proceedings}
}

@inproceedings{bhavya2020,
	title = {{Collective Development of Large Scale Data Science Products via Modularized Assignments: An Experience Report}},
	author = {Bhavya, Bhavya and Boughoula, Assma and Green, Aaron and Zhai, ChengXiang},
	year = {2020},
	month = {02},
	date = {2020-02-26},
	publisher = {Association for Computing Machinery},
	pages = {1200{\textendash}1206},
	booktitle = {SIGCSE '20},
	doi = {10.1145/3328778.3366961}
}

@article{hagen2020,
	title = {{Teaching Undergraduate Data Science for Information Schools}},
	author = {Hagen, Loni},
	year = {2020},
	month = {01},
	date = {2020-01-01},
	journal = {Education for Information},
	pages = {109--117},
	volume = {36},
	number = {2},
	doi = {10.3233/EFI-200372}
}

@inproceedings{fisler2022,
	title = {{Data-Centricity: Rethinking Introductory Computing to Support Data Science}},
	booktitle = {1st International Workshop on Data Systems Education},
	author = {Fisler, Kathi},
	year = {2022},
	month = {06},
	date = {2022-06-12},
	publisher = {Association for Computing Machinery},
	pages = {1{\textendash}3},
	series = {DataEd '22},
	doi = {10.1145/3531072.3535317}
}

@article{pieterman-bos2023,
	title = {{Integration of Philosophy of Science in Biomedical Data Science Education to Foster Better Scientific Practice}},
	author = {Pieterman-Bos, Annelies and van Mil, Marc H. W.},
	year = {2023},
	month = {12},
	date = {2023-12-01},
	journal = {Science and Education},
	pages = {1709--1738},
	volume = {32},
	number = {6},
	doi = {10.1007/s11191-022-00363-x}
}

@inproceedings{rao2019,
  title = {{Data Science Education Through Education Data: an End-to-End Perspective
}},
	booktitle = {2019 IEEE Integrated STEM Education Conference (ISEC)},
	author = {Rao, A. Ravishankar and Desai, Yashvi and Mishra, Kavita},
	year = {2019},
	pages = {300--307},
	doi = {10.1109/ISECon.2019.8881970},
}

@article{bilgin2022,
	title = {{Work Integrated Learning in Data Science and a Proposed Assessment Framework}},
	author = {Bilgin, Ayse Aysin Bombaci and Powell, Angela and Richards, Deborah},
	year = {2022},
	month = {07},
	date = {2022-07-04},
	journal = {Statistics Education Research Journal},
	pages = {12--12},
	volume = {21},
	number = {2},
	doi = {10.52041/serj.v21i2.26},
}

@inproceedings{haynes2019,
	title = {{Integrating Data Science into a General Education Information Technology Course: An Approach to Developing Data Savvy Undergraduates}},
	author = {Haynes, Malcolm and Groen, Joshua and Sturzinger, Eric and Zhu, Danny and Shafer, Justin and McGee, Timothy},
	year = {2019},
	month = {09},
	date = {2019-09-26},
	pages = {183{\textendash}188},
	booktitle = {SIGITE '19},
	doi = {10.1145/3349266.3351417}
}

@article{tucker2023,
	title = {{Teaching Statistics and Data Analysis with R}},
	author = {Tucker, Mary C. and Shaw, Stacy T. and Son, Ji Y. and Stigler, James W.},
	year = {2023},
	month = {01},
	date = {2023-01-02},
	journal = {Journal of Statistics and Data Science Education},
	pages = {18--32},
	volume = {31},
	number = {1},
	doi = {10.1080/26939169.2022.2089410}
}

@inproceedings{allen2021,
	title = {{Experiential Learning in Data Science: Developing an Interdisciplinary, Client-Sponsored Capstone Program}},
	author = {Allen, Genevera I.},
	year = {2021},
	month = {03},
	pages = {516{\textendash}522},
	booktitle = {SIGCSE '21}

}

@article{doudesis2022,
	title = {{Data Science in Undergraduate Medicine: Course Overview and Student Perspectives}},
	author = {Doudesis, Dimitrios and Manataki, Areti},
	year = {2022},
	month = {03},
	date = {2022-03},
	journal = {International Journal of Medical Informatics},
	pages = {104668},
	volume = {159},
	doi = {10.1016/j.ijmedinf.2021.104668},
	note = {PMID: 35033982},
	langid = {eng}
}

@inproceedings{blair2021,
	title = {{Establishing ABET Accreditation Criteria for Data Science}},
	author = {Blair, Jean R. S. and Jones, Lawrence and Leidig, Paul and Murray, Scott and Raj, Rajendra K. and Romanowski, Carol J.},
	year = {2021},
	month = {03},
	date = {2021-03-05},
	pages = {535{\textendash}540},
	booktitle = {SIGCSE '21}
}

@article{dill-mcfarland2021,
	title = {{An Integrated, Modular Approach to Data Science Education in Microbiology}},
	author = {Dill-McFarland, Kimberly A. and {König}, Stephan G. and Mazel, Florent and Oliver, David C. and McEwen, Lisa M. and Hong, Kris Y. and Hallam, Steven J.},
	year = {2021},
	month = {02},
	date = {2021-02-25},
	journal = {PLOS Computational Biology},
	pages = {e1008661},
	volume = {17},
	number = {2},
	doi = {10.1371/journal.pcbi.1008661},
}

@article{miah2020,
	title = {{A Design-Based Research Approach for Developing Data-Focussed Business Curricula}},
	author = {Miah, Shah J. and Solomonides, Ian and Gammack, John G.},
	year = {2020},
	month = {01},
	date = {2020-01-01},
	journal = {Education and Information Technologies},
	pages = {553--581},
	volume = {25},
	number = {1},
	doi = {10.1007/s10639-019-09981-5}
}

@article{yamamoto2021,
	title = {{Development of Online Learning Material for Data Science Programming Using 3D Puzzle}},
	author = {Yamamoto, Naoki and Ishida, Akio and Ogitsuka, Kazuki and Oishi, Nobuhiro and Murakami, Jun},
	year = {2021},
	date = {2021},
	journal = {International Journal of Information and Education Technology},
	pages = {154--163},
	volume = {11},
	number = {4},
	doi = {10.18178/ijiet.2021.11.4.1505}
}

@inproceedings{anslow2016,
	title = {{Datathons: An Experience Report of Data Hackathons for Data Science Education}},
	author = {Anslow, Craig and Brosz, John and Maurer, Frank and Boyes, Mike},
	year = {2016},
	month = {02},
	date = {2016-02-17},
	pages = {615{\textendash}620},
	booktitle = {SIGCSE '16}
}

@inproceedings{kross2019,
	title = {{Practitioners Teaching Data Science in Industry and Academia: Expectations, Workflows, and Challenges}},
	author = {Kross, Sean and Guo, Philip J.},
	year = {2019},
	month = {05},
	date = {2019-05-02},
	pages = {1{\textendash}14},
	booktitle = {CHI '19},
	doi = {10.1145/3290605.3300493}
}

@inproceedings{bilehassan2020,
  title={{A Comparative Study of the Academic Programs between Informatics BioInformatics and Data Science in the U.S}},
	author = {Bile Hassan, Ismail and Liu, Jigang},
	year = {2020},
	month = {07},
	booktitle = {2020 {IEEE} 44th Annual Computers, Software, and Applications Conference (COMPSAC)},
	pages = {165--171},
	doi = {10.1109/COMPSAC48688.2020.00030}
}

@inproceedings{kakeshita2022,
	title = {{Development of IPSJ Data Science Curriculum Standard}},
	author = {Kakeshita, Tetsuro and Ishii, Kazuo and Ishikawa, Yoshiharu and Matsubara, Hitoshi and Matsuo, Yutaka and Murata, Tsuyoshi and Nakano, Miyuki and Nakatani, Takako and Okumura, Haruhiko and Takahashi, Naoko and Takahashi, Norimitsu and Uchida, Gyo and Uematsu, Eriko and Saeki, Satoshi and Kato, Hiroshi},
	editor = {Passey, Don and Leahy, Denise and Williams, Lawrence and Holvikivi, Jaana and Ruohonen, Mikko},
	year = {2022},
	pages = {156--167},
	booktitle = {IFIP Advances in Information and Communication Technology},
	doi = {10.1007/978-3-030-97986-7_13}
}

@inproceedings{demchenko2019,
	title = {{Designing Customisable Data Science Curriculum Using Ontology for Data Science Competences and Body of Knowledge}},
	author = {Demchenko, Yuri and Comminiello, Luca and Reali, Gianluca},
	year = {2019},
	month = {03},
	pages = {124{\textendash}128},
	booktitle = {ICBDE '19},
	doi = {10.1145/3322134.3322143}
}

@article{betz2020,
	title = {{The Next Wave: We Will All Be Data Scientists}},
	author = {Betz, Margaret and Gundlach, Ellen and Hillery, Elizabett and Rickus, Jenna and Ward, Mark D.},
	year = {2020},
	date = {2020},
	journal = {Statistical Analysis and Data Mining: The ASA Data Science Journal},
	pages = {544--547},
	volume = {13},
	number = {6},
	doi = {https://doi.org/10.1002/sam.11476}
}

@inproceedings{banadaki2022,
  title = {{Enabling Data Science Education in {STEM} Disciplines through Supervised Undergraduate Research Experiences}},
	booktitle = {2022 ASEE Annual Conference and Exposition},
	author = {Banadaki, Yaser},
	year = {2022},
	date = {2022}
}

@article{song2017,
	title = {{Big Data and Data Science: Opportunities and Challenges of iSchools}},
	author = {Song, Il-Yeol and Zhu, Yongjun},
	year = {2017},
	month = {07},
	date = {2017-07-31},
	journal = {Journal of Data and Information Science},
	pages = {1--18},
	volume = {2},
	number = {3},
	doi = {10.1515/jdis-2017-0011}
}

@inproceedings{raj2019,
	title = {{An Empirical Approach to Understanding Data Science and Engineering Education}},
	author = {Raj, Rajendra K. and Parrish, Allen and Impagliazzo, John and Romanowski, Carol J. and Aly, Sherif G. and Bennett, Casey C. and Davis, Karen C. and McGettrick, Andrew and Pereira, Teresa Susana Mendes and Sundin, Lovisa},
	year = {2019},
	month = {12},
	date = {2019-12-18},
	pages = {73{\textendash}87},
	booktitle = {ITiCSE-WGR '19},
	doi = {10.1145/3344429.3372503}
}

@article{takemura2018,
	title = {{A New Era of Statistics and Data Science Education in Japanese Universities}},
	author = {Takemura, Akimichi},
	year = {2018},
	month = {06},
	date = {2018-06-01},
	journal = {Japanese Journal of Statistics and Data Science},
	pages = {109--116},
	volume = {1},
	number = {1},
	doi = {10.1007/s42081-018-0005-7}
}

@article{zakaria2023,
	title = {{Data Science Education Programmes in Middle Eastern Institutions: A Survey Study}},
	author = {Zakaria, Mahmoud Sherif},
	year = {2023},
	month = {03},
	date = {2023-03-01},
	journal = {IFLA Journal},
	pages = {157--179},
	volume = {49},
	number = {1},
	doi = {10.1177/03400352221113362}
}

@article{schmitt2023,
	title = {{Evaluation of EDISON's Data Science Competency Framework Through a Comparative Literature Analysis}},
	author = {Schmitt, Karl R. B. and Clark, Linda and Kinnaird, Katherine M. and Wertz, Ruth E. H. and Sandstede, {Björn}},
	year = {2023},
	month = {06},
	date = {2023-06-01},
	journal = {Foundations of Data Science},
	pages = {177--198},
	volume = {5},
	number = {2},
	doi = {10.3934/fods.2021031}
}

@inproceedings{wiktorski2016,
title={{Quantitative and Qualitative Analysis of Current Data Science Programs from Perspective of Data Science Competence Groups and Framework}},
	booktitle = {{2016 IEEE International Conference on Cloud Computing Technology and Science (CloudCom)}},
	author = {Wiktorski, Tomasz and Demchenko, Yuri and Belloum, Adam and Shirazi, Anoosheh},
	year = {2016},
	month = {12},
	date = {2016-12},
	pages = {633--638},
	doi = {10.1109/CloudCom.2016.0109},
	note = {ISSN: 2330-2186}
}

@article{oliver2021,
	title = {{Undergraduate Data Science Degrees Emphasize Computer Science and Statistics but Fall Short in Ethics Training and Domain-Specific Context}},
	author = {Oliver, Jeffrey C. and McNeil, Torbet},
	year = {2021},
	month = {03},
	date = {2021-03-25},
	journal = {PeerJ Computer Science},
	pages = {e441},
	volume = {7},
	doi = {10.7717/peerj-cs.441}
}

@article{ceccucci2015,
	title = {{The Effectiveness of Data Science as a means to achieve Proficiency in Scientific Literacy}},
	author = {Ceccucci, Wendy and Tamarkin, Dawn and Jones, Kiku},
	year = {2015},
	month = {07},
	date = {2015-07-01},
	journal = {Information Systems Education Journal},
	pages = {64},
	volume = {13},
	number = {4}
}

@article{song2016,
	title = {{Big Data and Data Science: What Should We Teach?}},
	author = {Song, Il-Yeol and Zhu, Yongjun},
	year = {2016},
	date = {2016},
	journal = {Expert Systems},
	pages = {364--373},
	volume = {33},
	number = {4},
	doi = {10.1111/exsy.12130}
}

@article{belloum2019,
	title = {{Bridging the Demand and the Offer in Data Science}},
	author = {Belloum, Adam S.Z. and Koulouzis, Spiros and Wiktorski, Tomasz and Manieri, Andrea},
	year = {2019},
	date = {2019},
	journal = {Concurrency and Computation: Practice and Experience},
	pages = {e5200},
	volume = {31},
	number = {17},
	doi = {10.1002/cpe.5200}
}

@inproceedings{davis2020,
  title = {{Ethics in Data Science Education}},
	booktitle = {2020 ASEE Virtual Annual Conference Content Access},
	author = {Davis, Karen C.},
	year = {2020},
	month = {06},
	date = {2020-06-22}
}

@article{shao2021,
	title = {{Exploring Potential Roles of Academic Libraries in Undergraduate Data Science Education Curriculum Development}},
	author = {Shao, Gang and Quintana, Jenny P. and Zakharov, Wei and Purzer, Senay and Kim, Eunhye},
	year = {2021},
	month = {03},
	date = {2021-03-01},
	journal = {The Journal of Academic Librarianship},
	pages = {102320},
	volume = {47},
	number = {2},
	doi = {10.1016/j.acalib.2021.102320}
}

@inproceedings{li2021,
	title = {{Exploring Interdisciplinary Data Science Education for Undergraduates: Preliminary Results}},
	booktitle = {International Conference on Information},
	author = {Li, Fanjie and Xiao, Zhiping and Ng, Jeremy Tzi Dong and Hu, Xiao},
	editor = {Toeppe, Katharina and Yan, Hui and Chu, Samuel Kai Wah},
	year = {2021},
	date = {2021},
	publisher = {Springer International Publishing},
	pages = {551--561},
	series = {Lecture Notes in Computer Science},
	doi = {10.1007/978-3-030-71292-1_43},
	address = {Cham},
	langid = {en}
}

@article{robeva2020,
	title = {{Changing the Nature of Quantitative Biology Education: Data Science as a Driver}},
	author = {Robeva, Raina S. and Jungck, John R. and Gross, Louis J.},
	year = {2020},
	month = {09},
	date = {2020-09-19},
	journal = {Bulletin of Mathematical Biology},
	pages = {127},
	volume = {82},
	number = {10},
	doi = {10.1007/s11538-020-00785-0}
}

@article{bonnell2022,
	title = {{Challenges and Issues in Data Science Education}},
	author = {Bonnell, Jerry and Ogihara, Mitsunori and Yesha, Yelena},
	year = {2022},
	date = {2022},
	journal = {Computer},
	pages = {63{\textendash}66},
	volume = {55},
	number = {2},
	note = {Publisher: IEEE}
}

@article{hazzan2021,
	title = {{A Journal for Interdisciplinary Data Science Education}},
	author = {Hazzan, Orit and Mike, Koby},
	year = {2021},
	date = {2021},
	journal = {Communications of the ACM},
	pages = {10{\textendash}11},
	volume = {64},
	number = {8},
	note = {Publisher: ACM New York, NY, USA}
}

@article{dinov2019,
	title = {{Quant Data Science meets Dexterous Artistry}},
	author = {Dinov, Ivo D.},
	year = {2019},
	month = {03},
	date = {2019-03},
	journal = {International Journal of Data Science and Analytics},
	pages = {81--86},
	volume = {7},
	number = {2},
	doi = {10.1007/s41060-018-0138-6},
	note = {PMID: 30923735
PMCID: PMC6433171},
	langid = {eng}
}

@article{engel2017,
	title = {{Statistical Literacy for Active Citizenship: A Call for Data Science Education}},
	author = {Engel, Joachim},
	year = {2017},
	date = {2017},
	journal = {Statistics Education Research Journal},
	pages = {44--49},
	volume = {16},
	number = {1},
	doi = {10.52041/serj.v16i1.213}
}

@book{nationalacademiesofsciences2018,
	title = {{Data Science for Undergraduates: Opportunities and Options}},
	author = {{National Academies of Sciences, Engineering and Medicine}},
	publisher ={{National Academies Press}},
	year = {2018},
	date = {2018}
}

@article{danyluk2021a,
	title = {Computing competencies for undergraduate data science curricula},
	author = {Danyluk, Andrea and Leidig, Paul and Bari, Anasse and Buck, Scott and Cassel, Lillian and Doyle, Maureen and Hines, Keegan and Ho, Tin Kam and McGettrick, Andrew and Qian, Weining and Schmitt, Karl and and Servin, Christian and Stefik, Andreas and Wang, Hongzhi},
	year = {2021},
	date = {2021}
}

@book{cooper1998syn,
  title={{Synthesizing Research: A Guide for Literature Reviews}},
  author={Cooper, Harris M},
  volume={2},
  year={1998},
  publisher={Sage}
}

@book{fraenkel2012design,
  title={{How to Design and Evaluate Research in Education (8th) ed.)}},
  author={Fraenkel, Jack and Wallen, Norman and Hyun, Helen},
  year={2012},
  publisher={McGraw-Hill}
}

@article{hicks2018,
	title = {{A Guide to Teaching Data Science}},
	author = {Hicks, Stephanie C. and Irizarry, Rafael A.},
	year = {2018},
	month = {10},
	date = {2018-10-02},
	journal = {The American Statistician},
	pages = {382--391},
	volume = {72},
	number = {4},
	doi = {10.1080/00031305.2017.1356747}
}

@incollection{hazzan2023,
  title={{What is Data Science?}},
  author={Hazzan, Orit and Mike, Koby},
  booktitle={Guide to Teaching Data Science: An Interdisciplinary Approach},
  pages={19--34},
  year={2023},
  publisher={Springer}
}

@article{cao2017,
  title={{Data Science: a Comprehensive Overview}},
  author={Cao, Longbing},
  journal={ACM Computing Surveys (CSUR)},
  volume={50},
  number={3},
  pages={1--42},
  year={2017},
  publisher={ACM New York, NY, USA}
}

@article{donoho2017,
	title = {{50 Years of Data Science}},
	author = {Donoho, David},
	year = {2017},
	month = {10},
	date = {2017-10-02},
	journal = {Journal of Computational and Graphical Statistics},
	pages = {745--766},
	volume = {26},
	number = {4},
	doi = {10.1080/10618600.2017.1384734}
}

@report{GAISE2016,
	title = {{Guidelines for Assessment and Instruction in Statistics Education {(GAISE)}}: College report},
	author = {GAISE},
	url = {http://www.amstat.org/education/gaise},
	year = {2016},
	misc = {Alexandria, VA: American Statistical Association}
}

@online{asacurricula,
author = {{American Statistical Association}},
title = {{Curriculum Guidelines for Undergraduate Programs in Statistical Science}},
year = {2014},
url={\url{https://www.amstat.org/docs/default-source/amstat-documents/edu-guidelines2014-11-15.pdf}},
urldate={2024-01-23}
}

@inproceedings{liu2022,
  title={{Prepare Data Science Program Student Outcomes and Curricula for ABET Accreditation}},
  author={Liu, David},
  booktitle={2022 ASEE Annual Conference \& Exposition},
  year={2022}
}

@misc{horton2022,
  title={{The Growing Importance of Reproducibility and Responsible Workflow in the Data Science and Statistics Curriculum}},
  author={Horton, Nicholas J and Alexander, Rohan and Parker, Micaela and Piekut, Aneta and Rundel, Colin},
  journal={Journal of Statistics and Data Science Education},
  volume={30},
  number={3},
  pages={207--208},
  year={2022},
  publisher={Taylor \& Francis}
}

@article{dogucu2022,
  title={{Tools and Recommendations for Reproducible Teaching}},
  author={Dogucu, Mine and {\c{C}}etinkaya-Rundel, Mine},
  journal={Journal of Statistics and Data Science Education},
  volume={30},
  number={3},
  pages={251--260},
  year={2022},
  publisher={Taylor \& Francis}
}

@article{punch2014introduction,
  title={{Introduction to Research Methods in Education}},
  author={Punch, Keith F and Oancea, Alis E},
  year={2014}
}

@inproceedings{cuadrado-gallego2021,
  title = {{Classification and Analysis of Techniques and Tools for Data Visualization Teaching}},
	booktitle = {{2021 IEEE Global Engineering Education Conference (EDUCON)}},
	author = {Cuadrado-Gallego, Juan J. and Demchenko, Yuri and Losada, Miguel A. and Ormandjieva, Olga},
	year = {2021},
	month = {04},
	date = {2021-04},
	pages = {1593--1599},
	doi = {10.1109/EDUCON46332.2021.9453917},
	url = {https://ieeexplore.ieee.org/abstract/document/9453917},
	note = {ISSN: 2165-9567}
}

@article{dogucu2024,
	title = {{Reproducibility in the Classroom}},
	author = {Dogucu, Mine},
	year = {2024},
	month = {10},
	date = {2024-10-09},
	doi = {10.1146/annurev-statistics-112723-034436},
	url = {https://www.annualreviews.org/content/journals/10.1146/annurev-statistics-112723-034436},
	note = {Publisher: Annual Reviews},
	langid = {en}
}

@article{baumer2022integrating,
  title={Integrating data science ethics into an undergraduate major: A case study},
  author={Baumer, Benjamin S and Garcia, Randi L and Kim, Albert Y and Kinnaird, Katherine M and Ott, Miles Q},
  journal={Journal of Statistics and Data Science Education},
  volume={30},
  number={1},
  pages={15--28},
  year={2022},
  publisher={Taylor \& Francis}
}

@article{finzer2013data,
  title={The data science education dilemma},
  author={Finzer, William},
  journal={Technology Innovations in Statistics Education},
  volume={7},
  number={2},
  year={2013}
}

@article{junk2020,
	title = {{Reproducibility and Replication of Experimental Particle Physics Results}},
	author = {Junk, Thomas R. and Lyons, Louis},
	year = {2020},
	month = {12},
	date = {2020-12-21},
	journal = {Harvard Data Science Review},
	volume = {2},
	number = {4},
	doi = {10.1162/99608f92.250f995b},
	url = {http://arxiv.org/abs/2009.06864},
	note = {arXiv:2009.06864 [physics]}
}

@article{chang,
	title = {{Is Economics Research Replicable? Sixty Published Papers from Thirteen Journals Say 'Usually Not'}},
	author = {Chang, Andrew C. and Li, Phillip},
	doi = {10.2139/ssrn.2669564},
	langid = {en},
	year = {2015},
	journal = {FEDS Working Paper No. 2015-083}
}

@article{opensciencecollaboration2015,
	title = {{Estimating the Reproducibility of Psychological Science}},
	author = {{Open Science Collaboration}},
	year = {2015},
	month = {08},
	date = {2015-08-28},
	journal = {Science},
	pages = {aac4716},
	volume = {349},
	number = {6251},
	doi = {10.1126/science.aac4716},
	url = {https://www.science.org/doi/10.1126/science.aac4716},
	note = {Publisher: American Association for the Advancement of Science}
}

@inbook{bausell2021,
	title = {{Publishing Issues and Their Impact on Reproducibility}},
	author = {Bausell, R. Barker},
	editor = {Bausell, R. Barker},
	year = {2021},
	month = {02},
	date = {2021-02-26},
	publisher = {Oxford University Press},
	pages = {0},
	doi = {10.1093/oso/9780197536537.003.0010},
	url = {https://doi.org/10.1093/oso/9780197536537.003.0010},
	note = {DOI: 10.1093/oso/9780197536537.003.0010}
}

@article{ioannidis2018,
	title = {{Thousands of Scientists Publish a Paper Every Five Days}},
	author = {Ioannidis, John P. A. and Klavans, Richard and Boyack, Kevin W.},
	year = {2018},
	month = {09},
	date = {2018-09},
	journal = {Nature},
	pages = {167--169},
	volume = {561},
	number = {7722},
	doi = {10.1038/d41586-018-06185-8},
	url = {https://www.nature.com/articles/d41586-018-06185-8},
	note = {Bandiera{\_}abtest: a
Cg{\_}type: Comment
Publisher: Nature Publishing Group
Subject{\_}term: Authorship, Publishing},
	langid = {en}
}

@online{nsfindicators,
author = {{National Science Foundation}},
title = {{Science and Engineering Indicators}},
year = {2018},
url={\url{https://www.nsf.gov/statistics/2018/nsb20181/assets/nsb20181.pdf}},
urldate={2024-12-27}
}

\end{document}